\documentclass[twocolumn]{revtex4}
\usepackage{graphicx}
\usepackage{dcolumn}
\usepackage{bm}
\usepackage{amsmath,amssymb}

\begin{document}

\title{Electrical conductance of atomic contacts in liquid environments}

\author{L. Gr\"{u}ter}
\author{M.T. Gonz\'{a}lez}
\author{R. Huber}
\author{M. Calame}
\email{michel.calame@unibas.ch}
\author{C. Sch{\"o}nenberger}
\affiliation{Institut f\"ur Physik, Universit\"at Basel,
Klingelbergstr.~82, CH-4056 Basel, Switzerland }
\date{\today}

\begin{abstract}

We present measurements of the electrical conductance $G$
at room temperature of mechanically controllable break junctions (MCBJ)
fabricated from Au in different solvents (octane, DCM, DMSO, and toluene) and
compare with measurements in air and vacuum.
In the high conductance regime \mbox{$G \agt G_0=2e^2/h$},
the environment plays a minor role, as
proven by the measured conductance histograms, which do not
depend on the environment. In contrast, the environment significantly
affects the electrical properties in the low conductance (tunneling)
regime \mbox{$G << G_0$}. Here, we observe a systematic and reproducible
lowering of the tunneling barrier height $\phi$. At shorter distances,
a transition to a strongly suppressed apparent barrier height
is observed in octane, providing evidence for the layering of solvent molecules
at small inter-electrodes separations.
The presented experimental configuration offers interesting
perspectives for the electrical characterization of single
molecules in a controlled environment.

\end{abstract}

\keywords{break junction, molecular electronics, atomic contacts}

\maketitle

% ------------------------------------------------------------------
% Main text
% ------------------------------------------------------------------

The perspectives to develop molecular devices based on single
molecule functionalities has recently attracted much attention
\cite{Nitzan03, Hush03}. Break junctions
\cite{Moreland85,Ruitenbeek96,Agrait03} experiments in vacuum
demonstrated that this technique is particularly suited to study
the electronic properties of single molecules
\cite{Reed97,Kergueris99,Smit02,Reichert03}. Adding a chemically
controllable environment in break junctions experiments would
offer a variety of interesting possibilities. The electronic
properties of a molecule can be tuned for instance by adjusting
its redox state using a liquid gate. A liquid environment offers
also a better control on the anchoring of the molecule to the
metallic constriction. The excellent efficiency of a liquid gate
has been demonstrated previously on carbon nanotube field-effect
transistors \cite{Krueger01}, while this effect has been recently
studied for organic molecules using a Scanning Tunneling
Microscope (STM) \cite{Xiao04}.

In this letter, we present a break junction setup with an integrated
liquid cell, allowing to explore the influence of solvents on
the electronic properties of atomic contacts.
The setup is used to study the variation of the electrical
conductance $G$ of Au junctions with their elongation in
the regime of tunneling (low conductance) and
true metallic contact (high conductance).
As solvents, we have used deionized water, dichloromethane (DCM),
dimethylsulfoxide (DMSO), octane and toluene. The last four are of
particular interest since they are potential organic solvents for
molecules relevant in molecular electronics. In addition, these
solvent cover a broad range of polarities. We also compare the
results with reference measurements obtained in vacuum and air.

\begin{figure}
  \includegraphics*[width=6cm]{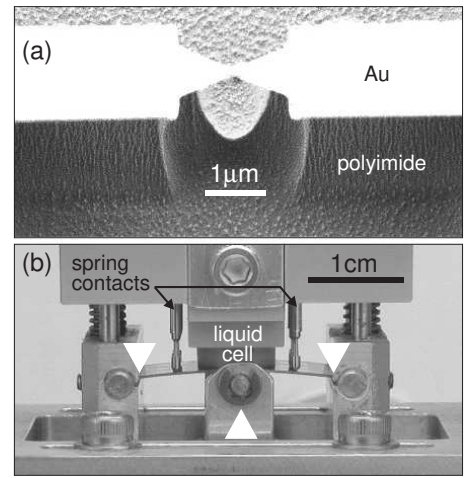} % two column
    \caption{\label{fig:setup}  (a) SEM image of one underetched Au
    junction defined by electron beam lithography. (b) View of the
    substrate mounted in the three point bending mechanism (white
    triangles) with a liquid cell pressed against the substrate.
  }
\end{figure}

The suspended metallic bridges were fabricated following a method
similar to the one described in Ref.~\cite{Ruitenbeek96}. We use
$0.3\times 10\times 24$~mm$^3$ polished phosphor bronze
substrates, spin-coated with a 2-3~$\mu$m thick polyimide layer
(PI2610, HD MicroSystems), and annealed for 30 min at
$200$~$^{\circ}$C in air, and 1 h at $400$~$^{\circ}$C and
in vacuum at $10^{-7}$~mbar. The metallic electrodes were produced with a
combination of UV and electron beam lithography, followed by a
Au-metallization step and an underetching of the junctions using
an oxygen plasma (80\% O$_2$ and 20\% CHF$_3$; 0.1~Torr; 100~W).
On a substrate, three junctions with a central constriction
80--150~nm wide (Fig.~\ref{fig:setup}(a)) are fabricated in
parallel. The resistance of the junctions ranges typically between
200~$\Omega$ and 300~$\Omega$ for 70~nm thick bridges.
The sample is mounted (unclamped) in the three-point-bending mechanism shown
in Fig.~\ref{fig:setup}(b). The distance between the
counter-supports is 20~mm. The vertical displacement of the
pushing-rod pressing on the sample from below is driven by a
stepper motor via a coupling gear, allowing for displacement
amplitudes up to a few millimeters with a resolution of 3~nm. To
perform measurements in a liquid, we integrated a liquid cell formed
by a portion of a $Viton^{\circledR}$ tube, enclosing a volume of
250~$\mu$l. The cell includes an inlet and outlet port allowing
the exchange of fluids in the course of the measurements. A tight
contact of the cell to the sample surface is ensured via a spring.
The electrical measurements were performed in a two-probe
configuration, using spring-loaded metallic tips.
A standard data acquisition board (National Instruments) was used both to
apply a constant bias voltage of \mbox{$0.1$\,V},
and to record the current in the junction as measured by a
current-voltage converter.

\begin{figure}
  \includegraphics*[width=7.5cm]{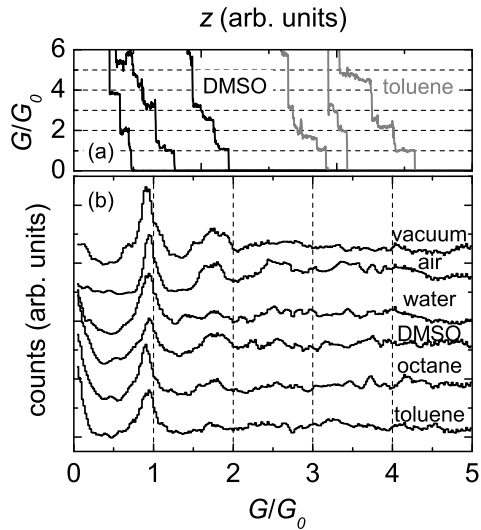} % two column
    \caption{\label{fig:histo} High conductance regime: (a)
    Electrical conductance $G$ scaled to $G_0=2e^2/h$ versus vertical displacement
    of the push-rod $z$, shifted horizontally for clarity, for one gold junction in
    DMSO and in toluene. (b) Conductance histograms, collected from
    many measurements of $G(z)$ of three junctions measured in different
    environments. The data for DMSO, octane and toluene correspond to
    one junction, those for air and water to a second one and those
    for vacuum to a third one. The histograms are shifted
    vertically for clarity.
  }
\end{figure}

In the high conductance regime ($G \agt G_0=2e^2/h$),
for each environment, we collected
$100-130$ curves of the decrease of the junction conductance as a
function of the vertical displacement $z$ of the pushing-rod while
extending and, eventually, breaking the constriction. Typical
single conductance curves are shown
in Fig.~\ref{fig:histo}(a), whereas Fig.~\ref{fig:histo}(b) displays
the conductance histogram over all measured curves.
In agreement with previous work in air, vacuum and at low temperatures (He),
there are large junction-to-junction variations in $G(z)$.
Individual curves display plateaux at different conductance values.
The clearest and most reproducible plateau is the one close to $1\,G_0$,
as expected for monovalent wires \cite{Scheer98}, and gives rise
to a clear peak in the histograms shown in
Figure~\ref{fig:histo}(b). Each histogram is built from all the
conductance traces measured in one environment (bin width:
0.02~$G_0$) and normalized by the total number of counts. Because
no striking difference between histograms for different environments are apparent,
we conclude that the environment plays a minor rule in
the electronic properties of atomic contacts in the {\em high} conductance regime.

\begin{figure}
  \includegraphics*[width=7.5cm]{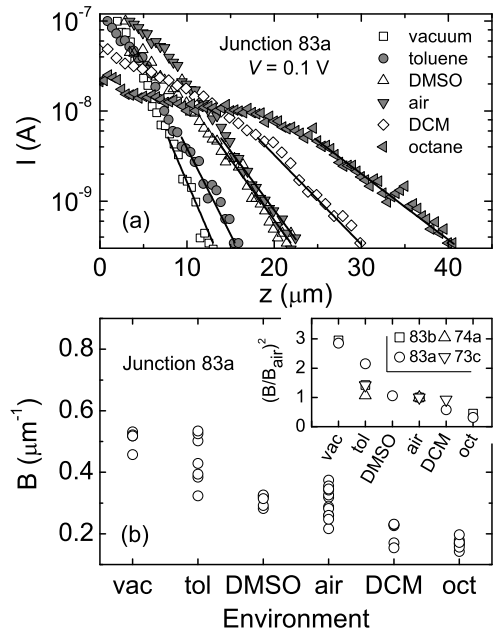} % two column
    \caption{\label{fig:tunnel} Tunneling regime: (a) Raw data for the
    electrical current $I$ versus vertical displacement of the push-rod $z$ for a
    single gold junction (83a) in different environments.
    The curves are shifted horizontally for clarity, and the zero in $z$
    has been arbitrarily chosen. The solid lines are obtained from fits
    of $I$ to an exponential law, i.e. $ln(I) = -B I + const$.
    (b) Results for the slope $B$ from fitting an exponential law to
    several sets of curves similar to that shown in part (a). The
    ratio of the mean value of $B$ in different environments
    with respect to that of air has been plotted in the inset for four
    different junctions.
  }
\end{figure}

We now focus on the behavior of the junctions in the tunneling
regime ($G$ between $4\times 10^{-5}$ and $0.01~G_0=2e^2/h$).
Several sets of $10$ consecutive open-close cycles (up to a
conductance of a few $G_0$) were recorded for the same junction in
different environments. Fig.~\ref{fig:tunnel}(a) shows
representative raw curves obtained when the junction was being closed.
% which is the relevant process to capture molecules within a break junction.
In the low current regime ($\leq5\times10^{-9}$~A), subsequent curves show little variation,
and exhibit a well defined linear regime in a semi-log
representation. At higher currents, we observed larger
curve-to-curve variations that can be attributed to differences in
the microscopic configurations of the junctions. Whenever
measuring in a medium different from vacuum, a clear decrease of
the slope at low currents is observed which is characteristic for
each environment. We emphasize that the slope is not correlated with
the order in which the environments were tested and, hence, with
a possible accumulation of contamination in the junction.
% For the junction of Fig.~\ref{fig:tunnel}(a), the measurement sequence
% was: vacuum, octane, air, toluene, DCM, and DMSO.

We consider the simple expression for the tunneling current at low
bias voltage through a square barrier of height $\phi$ and
thickness $d$, $I \propto \exp[-2 d \sqrt{2 m \phi}/\hbar$], where
$m$ is the electron mass and $d=r(z-z_0)$ the distance between
the two extremities  of the Au contacts on either side.
Here, $r$ is the mechanical reduction factor, which can be
obtained from the geometry of the junction \cite{Agrait03,Vrouwe??}
and $z_0$ is a constant.
We then have  $\ln(I) = -Bz + const$, where $B=2r\sqrt{2m\phi}/\hbar$
$= 1.025 [\mathrm{eV}^{-0.5}\mathrm{\AA}^{-1}]\sqrt{\phi} r$. Hence, the slope
of $ln(I)$ vs $-z$ is given by $B \propto \sqrt{\phi}$. A large (small) slope
corresponds to a large (small) barrier height $\phi$.
To reduce the effect of fluctuations, the slope $B$ was deduced in the
low current regime using all $10$ curves of each set of measurements simultaneously.
% imposing as a fitting restriction a common slope $B$ for all of the curves.
The result of several sets is shown in Fig.~\ref{fig:tunnel}(b) for the same
junction as in Fig.~\ref{fig:tunnel}(a). The lines in Fig.~\ref{fig:tunnel}(a)
correspond to the averaged values of $B$ for each environment.

We performed the same analysis for three additional samples and found
differences up to a factor of $3$ in $B$ for a given environment.
This large scattering cannot be assigned to $\phi$, because
unphysically large barrier heights would then results. The scattering
is due to sample-to-sample variation in the mechanical reduction factor $r$.
This is proven with the inset of Fig.~\ref{fig:tunnel}(b) which
illustrates the ratio $(B/B_{air})^2$ for each environment and
for four {\em different} junctions. This ratio is directly proportional to
the ratio of the effective barrier heights and does {\em not} depend on $r$.
We conclude, that $\phi$ does depend on the environment in a specific and
reproducible way.

Reduced effective tunneling barrier heights $\phi$ have been
consistently observed in STM experiments performed in aqueous
environments \cite{Schmickler96,Lindsay93,Tao00}. Although a
detailed model providing a quantitative account for this decrease
is still lacking, we point out that the observed decrease in $\phi$
does not show a simple correlation with the dielectric constant
(whether static or optical) of the medium in between the
electrodes~\cite{Schmickler90,Sebastian93}.

As we have stressed before, the mechanical reduction factor
$r$ presents a large uncertainty. Based on the actual geometry, we estimate
$r_g=6tu/L^2\simeq5\times10^{-6}$ \cite{Ruitenbeek96,Young89}.
In contrast, if we use the measured $B$ for vacuum and
fix $\phi$ to the established value of
\mbox{$3.5-5$\,eV} \cite{Gimzewski87},
we get values of $r\simeq5\times10^{-5}$, one
order of magnitude higher than $r_g$.

The cause for this discrepancy is likely caused by plastic deformation of
the substrate and flow of the polyimide layer in the vicinity of the junction
due to inhomogeneous stress \cite{Agrait03,Vrouwe??}. That plastic deformation
has to be considered is confirmed by the presence of a residual bending
of the substrate after several loading and unloading cycles in some experiments.
Moreover, we estimate that a vertical displacement $z$ of \mbox{$\sim 1$\,mm},
which is a typical excursion needed to break the junction, is
sufficient to reach the tensile strength of phosphor bronze.
Plastic deformation changes the bending geometry and can
substantially lower the mechanical reduction factor, as observed
in our experiment.

In addition to the dependence of $\phi$ on the environment, there is
a striking change of the slope $B$ to smaller values at small $z$ values,
observed in the octane curve [$\vartriangleleft$, Fig.~\ref{fig:tunnel}(a)]
and to a lesser extend in the DCM curve ($\lozenge$).
This behavior could be reliably observed for all curves measured
in octane after a number of open--close cycles ($>30$), allowing the junction to
stabilize. The observed lowering of the apparent barrier height
$\phi$ is likely caused by the discrete (molecular) nature of the liquid and its behavior
in confined geometries \cite{Bhushan95,Persson04}.
STM investigations have shown that n-alkane chains tend to
self-organize parallel to a Au(111) surface \cite{Marchenko02}.
A layering of the octane molecules at
short inter-electrodes separations can explain the reduced
slope of $ln(I)$ vs $z$ at small gap sizes: for a given vertical
displacement $z$, the effective shortening of the gap becomes smaller
due to the mechanical resistance opposed by the layered octane
molecules in between the Au electrodes, deforming the Au extremities.
Consequently, a more gentle current increase with vertical displacement
results.

% We suspect that mechanical effects linked to the
% discrete nature of the dielectric medium within the electrodes gap
% can also account for the variation of $B$ with the environment in
% the low current region.

In conclusion, we have shown that micro-fabricated break junctions
can be operated in a liquid environment opening interesting
perspectives to study single-molecule devices in a chemically
active environment. We observed that the solvent significantly
affects the tunneling regime and obtained good evidence
that the confinement of the liquid can lead to molecular layering
effects within the gap. We emphasize that the strain limitation of
the substrate causes significant variations in the mechanical
reduction factor. Despite this difficulty, we could observe a
systematic and reproducible trend in the variation of the
tunneling barrier height with different solvents.

We are indebted to H. Breitenstein, S. Jakob, P. Reimann and M.
Steinacher for technical support. We also thank C. Urbina for
kindly hosting one of us (L.G.), and A. Baratoff for helpful
discussions. This work benefitted from the support of the Swiss
National Center of Competence in Research ``Nanoscale Science''
and the European Science Foundation through the Eurocore's programme
on Self-Organized Nanostructures (SONS).

%---------------------------------------------------------------------------
% End of Main text
%---------------------------------------------------------------------------
\bibliographystyle{aip}

\begin{thebibliography}{99}

\bibitem{Nitzan03}
A. Nitzan and M.A. Ratner, Science \textbf{300}, 1384 (2003).

\bibitem{Hush03}
N. S. Hush, Ann. N.Y. Acad. Sci. \textbf{1006}, 1 (2003).

\bibitem{Moreland85}
J. Moreland, J.W. Ekin, J. Appl. Phys. \textbf{3888}, 58 (1985).

\bibitem{Ruitenbeek96}
J.M. van Ruitenbeek, A. Alvarez, I. Pi\~{n}eyro, C. Grahmann, P.
Joyez, M.H. Devoret, D. Esteve, C. Urbina,  Rev. Sci. Instrum.
\textbf{67}, 108 (1996).

\bibitem{Agrait03}
N. Agra\"{\i}t, A. Levy Yeyati, J. M. van Ruitenbeek, Physics Reports
\textbf{377}, 81 (2003).

\bibitem{Reed97}
M.A. Reed, C. Zhou, C.J. Muller, T.P. Burgin, J.M. Tour, Science
\textbf{278}, 252 (1997).

\bibitem{Kergueris99}
C. Kergueris, J.-P. Bourgoin, S. Palacin, D. Esteve, C. Urbina, M.
Magoga, C. Joachim, Phys. Rev. B \textbf{59}, 12505 (1999).

\bibitem{Smit02}
R. H. M. Smit, Y. Noat, C. Untiedt, N.D. Lang, M. van Hemert, J.M.
van Ruitenbeek,  Nature \textbf{906}, 419 (2002)

\bibitem{Reichert03}
J. Reichert, H.B. Weber, M. Mayor, H. V. L\"{o}hneysen, Appl. Phys.
Lett. \textbf{82}, 4137 (2003).

\bibitem{Krueger01}
M. Kr\"{u}ger, M.R. Buitelaar, T. Nussbaumer, C. Sch\"{o}nenberger,
Appl. Phys. Lett. \textbf{78}, 1291 (2001).

\bibitem{Xiao04}
X. Xiao, B. Xu, N. J. Tao, Nano Lett. \textbf{4}, 267 (2004).

\bibitem{Vrouwe??}
S. A. G. Vrouwe, E. van der Giessen, S. J. van der Molen, D. Dulic, M. L. Trouwborst, and
B. J. van Wees, preprint.

\bibitem{Scheer98}
E. Scheer, N. Agra\"{\i}t, J.C. Cuevas, A. Levy Yeyati, B. Ludolph,
A. Martin-Rodero, G. Rubio Bollinger, J.M. van Ruitenbeek, C. Urbina,
Nature \textbf{394}, 154 (1998).

\bibitem{Schmickler96}
W. Schmickler, Chem. Rev. \textbf{96}, 3177 (1996).

\bibitem{Lindsay93}
S. M. Lindsay, T. W. Jing, J. Pan, D. Lampner, A. Vaught, J. P.
Lewis, O. F. Sankey in {\it Nanoscale Probes of the Solid/Liquid
Interface, NATO ASI Series E} \textbf{1993}, {\it Vol 288}, 25.

\bibitem{Tao00}
N. J. Tao, C. Z. Li, H. X. He, J. Electroanal. Chem. \textbf{492}, 81
(2000).

\bibitem{Schmickler90}
W. Schmickler, D. J. Henderson, J. Electroanal. Chem. \textbf{290},
283 (1990).

\bibitem{Sebastian93}
K.L Sebastian, G. Doyen,  J. Chem. Phys. \textbf{99}, 6677 (1993).

\bibitem{Young89}
Calculated using the equations for a simply supported beam. See
e.g. W. C. Young, {\it Roark's Formulas for Stress and strain}
(McGraw-Hill, New York, 1989), Chap. 7.

\bibitem{Gimzewski87}
J. K. Gimzewski, R. M\"{o}ller, Phys. Rev. B \textbf{36}, 1284
(1987).

\bibitem{Bhushan95}
B. Bhushan, J. N. Israelachvili, U. Landman, Nature \textbf{374},
607 (1995).

\bibitem{Persson04}
B. N. J. Persson, F. Mugele, J. Phys.: Condens. Matter \textbf{16}
R295 (2004).

\bibitem{Marchenko02}
A. Marchenko, S. Lukyanets, J. Cousty, Phys. Rev. B \textbf{65},
045414 (2002).

\end{thebibliography}

%-------------------------------------------------------------
\end{document}